\documentclass[runningheads]{llncs}
\usepackage[T1]{fontenc}
\usepackage{dsfont}
\usepackage{graphicx,verbatim}
\usepackage{amsmath,amssymb}
\usepackage{booktabs}
\usepackage{siunitx}
\begin{document}
\title{Parameter-Efficient Adaptation of SAM~3 for Automated ITV Generation from 4DCT Images}
\titlerunning{SAM~3 Adaptation for Automated ITV Generation from 4DCT}

\author{Changwoo Song\inst{1,2}}
\authorrunning{C. Song}
\institute{Oncosoft Inc., Seoul, South Korea \and
Department of Computer Science \& Engineering, Chungnam National University, Daejeon, South Korea\\
\email{song921216@gmail.com}}

\maketitle         
\begin{abstract}
Four-dimensional computed tomography (4DCT) captures the full respiratory cycle of thoracic anatomy, yet current Internal Target Volume contouring workflows process each phase in isolation, discarding temporal coherence and leaving contours vulnerable to phase-specific artifacts.
We present a lightweight framework that applies parameter-efficient fine-tuning to the Segment Anything Model~3 (SAM~3) via low-rank adaptation (LoRA) to align its text-prompted segmentation with the medical domain using only seven annotated 3D CT volumes.
Furthermore, the framework incorporates a hard negative mining strategy to improve boundary discrimination in low-contrast thoracic regions.
At inference, phase-wise predictions are refined through phase-coherent temporal filtering and spatial connectivity analysis.
Since respiratory motion is continuous and periodic, genuine anatomy appears in contiguous blocks of phases, whereas transient artifacts appear sporadically and are thus effectively suppressed.
Experiments on pulmonary and cardiac structures yield median Dice scores of 0.968 and 0.910 with 95th-percentile Hausdorff distances of 0.998\,mm and 2.931\,mm, respectively.
The proposed framework effectively eliminates the severe false-positive predictions inherent in the zero-shot inference of the unadapted SAM~3.
With only seven annotated volumes, the framework retains over 95\% of full-data accuracy, and the entire pipeline is trainable on a single consumer-grade GPU, demonstrating a scalable, data-efficient solution for adaptive radiotherapy.


\end{abstract}

\section{Introduction}

Respiratory motion remains a principal source of geometric uncertainty in thoracic radiotherapy.
The Internal Target Volume (ITV), defined as the envelope of the clinical target volume across all respiratory phases~\cite{icru1999}, is the standard approach for encapsulating this motion~\cite{underberg2005use,keall2006management}.
Four-dimensional computed tomography (4DCT)~\cite{vedam2003acquiring} provides the temporal resolution needed to capture the full breathing cycle; however, current clinical practice typically derives the ITV by contouring each phase independently and merging the results, a labor-intensive procedure that scales poorly with the number of structures and phases~\cite{keall2006management,cardenas2019advances}.

Deep learning has shown remarkable success in automating medical image segmentation~\cite{ronneberger2015u,isensee2021nnu,hatamizadeh2021unetrtransformers3dmedical,wasserthal2023totalsegmentator}, yet deploying such models in radiotherapy workflows faces persistent obstacles: large annotation requirements~\cite{ma2024segment}---especially for 4DCT with ten or more phase volumes per scan---prohibitive GPU costs, and the neglect of inter-phase temporal coherence that could suppress transient artifacts.

Foundation models such as the Segment Anything Model (SAM)~\cite{kirillov2023segment,ravi2024sam} have demonstrated strong zero-shot generalization on natural images, yet the substantial domain gap to CT---low soft-tissue contrast, narrow Hounsfield-unit windows, and absence of color cues---renders direct application unreliable~\cite{ma2024segment,zhu2024medicalsam2segment,wu2023medicalsamadapteradapting}.
Full fine-tuning of a billion-parameter model demands hundreds of annotated volumes and multi-GPU infrastructure, placing it out of reach for most clinical sites.
Parameter-Efficient Fine-Tuning (PEFT) via Low-Rank Adaptation (LoRA)~\cite{hu2021loralowrankadaptationlarge} resolves this tension: by injecting small trainable rank-decomposition matrices into frozen transformer layers, LoRA enables effective domain transfer with minimal data and a single consumer-grade GPU~\cite{ding2023parameter}.

In this work, we present a framework that bridges the gap between foundation-model capability and clinical 4DCT requirements.
Our contributions are threefold:
\begin{enumerate}
    \item We demonstrate that text-prompted, LoRA-based PEFT of SAM~3 achieves robust thoracic segmentation with high data efficiency---requiring only seven 3D CT volumes. This serves as a highly efficient proof-of-concept that generalizes directly to unseen 4DCT phases in a zero-shot manner, incurring less than a 0.04 DSC drop on 3D CT benchmarks compared to 79-case training.
    \item We introduce a hard negative mining (HNM) strategy tailored to low-contrast thoracic boundaries, improving boundary discrimination without additional annotation effort.
    \item We propose a spatiotemporal filtering strategy leveraging respiratory periodicity to retain only voxels in contiguous phase blocks. This robustly suppresses sporadic artifacts inherent in zero-shot predictions, yielding a clinically plausible ITV representation verified by a radiation oncologist.
\end{enumerate}

\section{Methods}

\subsection{Overview}

\begin{figure}[t]
\centering
\includegraphics[width=\textwidth]{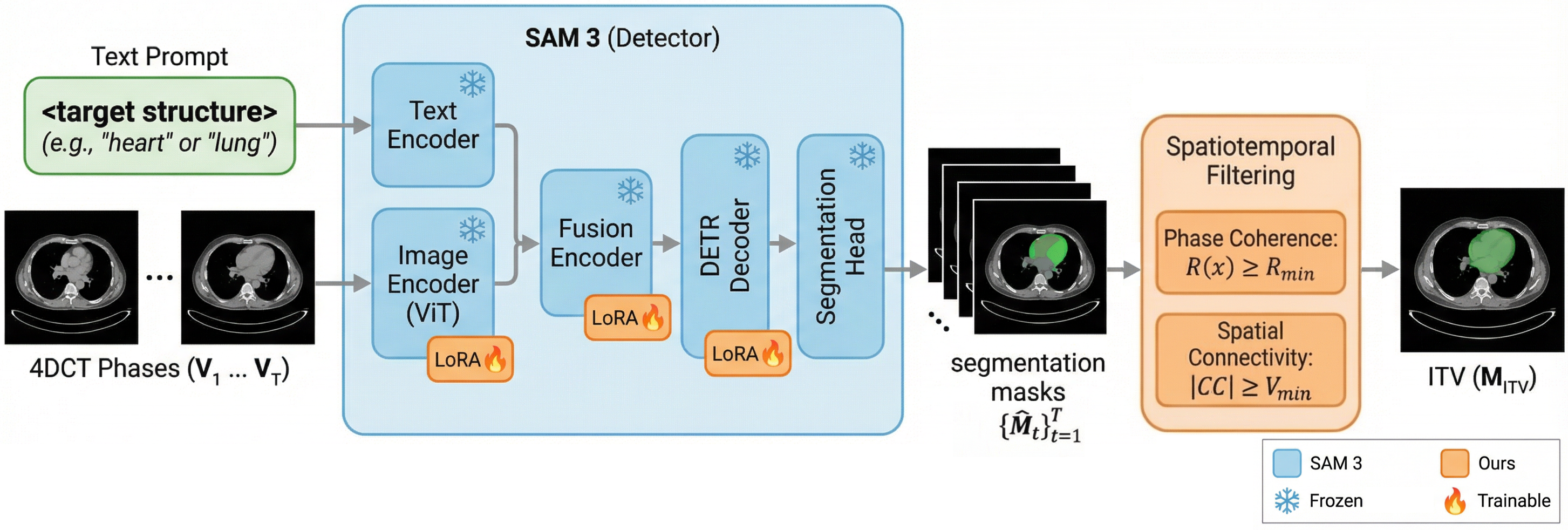}
\caption{Overview of the proposed framework. SAM~3 is adapted via LoRA and prompted with target structure names. Phase-wise 3D predictions are filtered by respiratory phase coherence and spatial connectivity to produce a unified ITV.}
\label{fig:pipeline}
\end{figure}

Given a 4DCT acquisition $\mathcal{V} = \{V_1, V_2, \ldots, V_T\}$ consisting of $T$ respiratory phase volumes (typically $T{=}10$), our goal is to produce a consolidated ITV mask $M_{\text{ITV}}$ for each target structure.
The pipeline comprises three key components: (1)~LoRA-adapted SAM~3 generates per-phase segmentation masks $\{\hat{M}_t\}_{t=1}^{T}$ using text prompts, (2)~an HNM strategy applied during training to refine the learning signal in ambiguous regions, and (3)~a spatiotemporal filter used at inference to fuse the $T$ predictions into a single ITV.
The overall inference architecture is illustrated in Fig.~\ref{fig:pipeline}.

\subsection{Parameter-Efficient Fine-Tuning with LoRA}

SAM~3~\cite{carion2025sam3segmentconcepts}, the latest iteration of the Segment Anything Model, employs a vision transformer (ViT)~\cite{dosovitskiy2021imageworth16x16words} backbone pretrained on large-scale natural image--video datasets.
While its architecture generalizes well to diverse visual domains, the distribution gap between natural scenes and low-contrast CT imagery necessitates domain adaptation.

Full fine-tuning of the entire model is prohibitive for clinical sites with limited GPU resources; we therefore freeze all pretrained weights and inject LoRA modules into all attention projection matrices (query, key, value, and output) of each transformer block.
LoRA parameterizes the update of a frozen weight matrix $W_0 \in \mathbb{R}^{d \times k}$ via a low-rank decomposition $W = W_0 + BA$, where $B \in \mathbb{R}^{d \times r}$, $A \in \mathbb{R}^{r \times k}$, and rank $r \ll \min(d, k)$~\cite{hu2021loralowrankadaptationlarge}.
Only $A$ and $B$ are trained, reducing the number of trainable parameters by over two orders of magnitude compared to full fine-tuning.

SAM~3 accepts multi-modal prompts; we employ text prompts specifying the target structure name (e.g., ``heart'', ``lung'') to condition the decoder, eliminating the need for interactive point or bounding-box annotations during inference.

\subsection{Hard Negative Mining}

Thoracic CT exhibits regions where tissue boundaries are inherently ambiguous---for instance, the interface between the heart and the adjacent mediastinal fat, or the lung boundary near the diaphragm.
Standard training with uniform sampling under-represents these challenging voxels, leading to false-positive predictions that propagate into the ITV.

Inspired by contrastive learning and hard example mining~\cite{robinson2021contrastivelearninghardnegative,lin2017focal}, we integrate an HNM strategy into our loss objective.
After each forward pass, we identify voxels that are predicted as foreground but labeled as background (i.e., false positives), and rank them by prediction confidence.
These voxels are re-weighted in the loss computation, which combines a Dice loss~\cite{milletari2016vnetfullyconvolutionalneural} with a focused binary cross-entropy term:
\begin{equation}
    \mathcal{L} = \mathcal{L}_{\text{Dice}} + \lambda \cdot \frac{1}{|\mathcal{H}|}\sum_{i \in \mathcal{H}} \ell_{\text{BCE}}(\hat{p}_i, y_i),
\label{eq:loss}
\end{equation}
where $\mathcal{H}$ denotes the hard negative set, $\hat{p}_i$ is the predicted probability, $y_i$ is the ground-truth label, and $\lambda$ controls the mining strength.
This encourages the model to allocate representational capacity to the most confusable regions, directly reducing false-positive rates without requiring additional annotations.

\subsection{Spatiotemporal Filtering for ITV Generation}

A key observation is that anatomical structures move along a continuous, quasi-periodic trajectory throughout the respiratory cycle.
Consequently, if a voxel truly belongs to a target organ, it should be detected in a contiguous block of respiratory phases rather than sporadically.
Transient artifacts---such as motion blur at phase transitions or streak artifacts---appear in isolated, non-adjacent phases and can therefore be distinguished from genuine anatomy.
We exploit this property through a two-stage filtering process.

\paragraph{Phase Coherence.}
For each voxel position $\mathbf{x}$, let $\hat{M}_t(\mathbf{x}) \in \{0,1\}$ denote the binary prediction from phase~$t$.
Since the respiratory cycle is periodic (phase~$T$ is adjacent to phase~$1$), we define the \emph{maximum circular run length} $R(\mathbf{x})$ as the longest contiguous block of consecutive phases in which the voxel is detected.
A voxel is retained if $R(\mathbf{x}) \geq R_{\min}$, where $R_{\min}$ is a minimum-consecutive-phase threshold.

\paragraph{Spatial Connectivity.}
After phase-coherence filtering, residual isolated voxel clusters may persist due to prediction noise.
We apply connected-component analysis and discard components smaller than a volume threshold $V_{\min}$, retaining only spatially coherent regions.
The final ITV is obtained as:
\begin{equation}
    M_{\text{ITV}}(\mathbf{x}) = \text{CC}_{\geq V_{\min}}\!\bigl[\mathds{1}[R(\mathbf{x}) \geq R_{\min}]\bigr],
\label{eq:itv}
\end{equation}
where $\text{CC}_{\geq V_{\min}}[\cdot]$ retains connected components exceeding $V_{\min}$ voxels.

\section{Experiments and Results}

\subsection{Dataset and Implementation Details}

\paragraph{Dataset.}
Two distinct datasets were used.
For \textit{LoRA fine-tuning}, we curated a separate collection of conventional 3D CT volumes with expert-delineated heart and bilateral lung contours.
The full set of 89 cases was split into 79 for training, 5 for validation, and 5 for testing.
We investigated training set sizes of 79, 39, 19, and 7 cases (100\%, 50\%, 25\%, and 10\% of the training set) to evaluate data efficiency; the minimal configuration uses only seven annotated 3D CT volumes.
Training slices were augmented with boundary-proximal background slices---axial slices adjacent to organ boundaries that contain no foreground---at a ratio of 30\% relative to positive slices.
For \textit{4DCT evaluation}, we adopted the publicly available "CT-vs-PET-Ventilation-Imaging" dataset from The Cancer Imaging Archive (TCIA)~\cite{https://doi.org/10.7937/3ppx-7s22}, comprising 4DCT scans of 20 lung cancer patients sorted into 10 respiratory phase bins (0\%--90\%).
The LoRA-adapted model, trained exclusively on 3D CT, was applied to each phase without further fine-tuning.

\paragraph{Implementation.}
Experiments were conducted on a single NVIDIA RTX~3080 GPU (\SI{10}{GB} VRAM).
SAM~3 was initialized from its publicly available pretrained weights.
LoRA modules ($r{=}8$) were injected into all attention projections of every transformer layer, yielding approximately \SI{1.59}{M} trainable parameters---less than 0.19\% of the \SI{842}{M} total.
While prior medical SAM adaptations~\cite{zhang2023customizedsegmentmodelmedical} often restrict LoRA to specific projections to minimize overhead, we extended it to the entire attention block to fully leverage SAM~3's representation power.
We trained for up to 20 epochs using the AdamW optimizer~\cite{loshchilov2019decoupledweightdecayregularization} with a fixed learning rate of $1 \times 10^{-4}$ and early stopping (patience of 5 epochs on validation DSC).
The HNM ratio was set to $|\mathcal{H}|{=}256$ voxels per batch, with $\lambda{=}0.5$.
To prevent training collapse from premature boundary penalization, $\lambda$ is held at zero for the first three epochs (warm-up) and then linearly increased to its target value over five epochs.
For spatiotemporal filtering, we set $R_{\min}{=}3$ (requiring detection in at least three consecutive respiratory phases) and $V_{\min}{=}100$ voxels.

\subsection{Quantitative Results}

\begin{figure*}[!t]
\centering
\includegraphics[width=\textwidth]{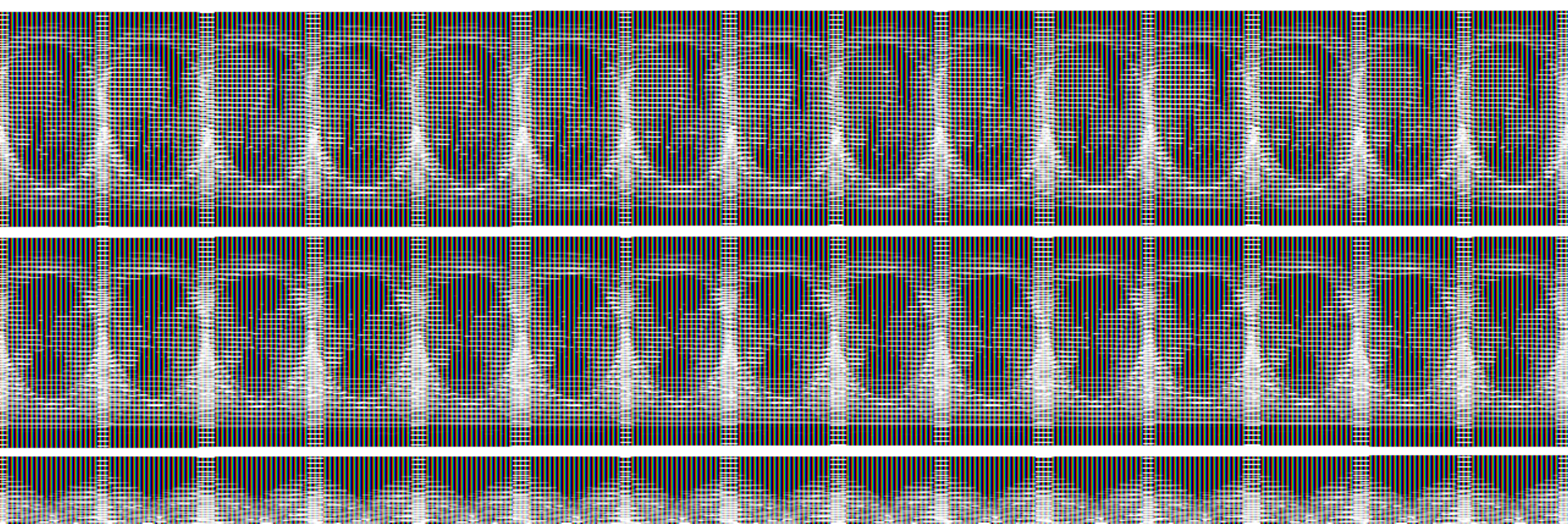}
\caption{Qualitative comparison across three axial slices. Red: prediction; green: ground truth. \textbf{(a)}~Input CT. \textbf{(b)}~Zero-shot SAM~3. \textbf{(c)}~LoRA. \textbf{(d)}~LoRA~+~HNM. \textbf{(e)}~Ground truth. Top to bottom: central, boundary, and distant cardiac slices.}
\label{fig:qualitative}
\end{figure*}

Without adaptation, zero-shot SAM~3 yields near-zero DSC (0.028 and 0.082 for the ``heart'' and ``lung'' prompts, respectively) on thoracic CT, confirming the large domain gap between natural and medical images.
LoRA adaptation improves segmentation accuracy (DSC of 0.603 and 0.663, respectively); however, the 95th-percentile Hausdorff Distance (HD95) of 111.8\,mm and 89.6\,mm indicate persistent distant false positives outside the target organs.
Incorporating HNM resolves this issue: the heart DSC reaches 0.910 with an HD95 of 2.931\,mm, and the lung DSC reaches 0.968 with an HD95 of 0.998\,mm, as summarized in Table~\ref{tab:main}.
This effect is illustrated qualitatively in Fig.~\ref{fig:qualitative}: at central slices both adapted models approximate the ground truth, whereas at boundary and distant slices only the model trained with HNM effectively suppresses false positives.

\begin{table}[t]
\centering
\caption{Segmentation performance comparison on the held-out test set. Best results are \underline{underlined}.}
\label{tab:main}
\begin{tabular}{l cc cc}
\toprule
 & \multicolumn{2}{c}{\textbf{``heart''}} & \multicolumn{2}{c}{\textbf{``lung''}} \\
\cmidrule(lr){2-3} \cmidrule(lr){4-5}
\textbf{Method} & DSC & HD95 & DSC & HD95 \\
\midrule
SAM~3 (zero-shot) & 0.028 & 319.452 & 0.082 & 328.962 \\
SAM~3 + LoRA & 0.603 & 111.828 & 0.663 & 89.608 \\
SAM~3 + LoRA + HNM & \underline{0.910} & \underline{2.931} & \underline{0.968} & \underline{0.998} \\
\bottomrule
\end{tabular}
\end{table}

\subsection{Ablation Studies}

\begin{figure}[t]
\centering
\includegraphics[width=\columnwidth]{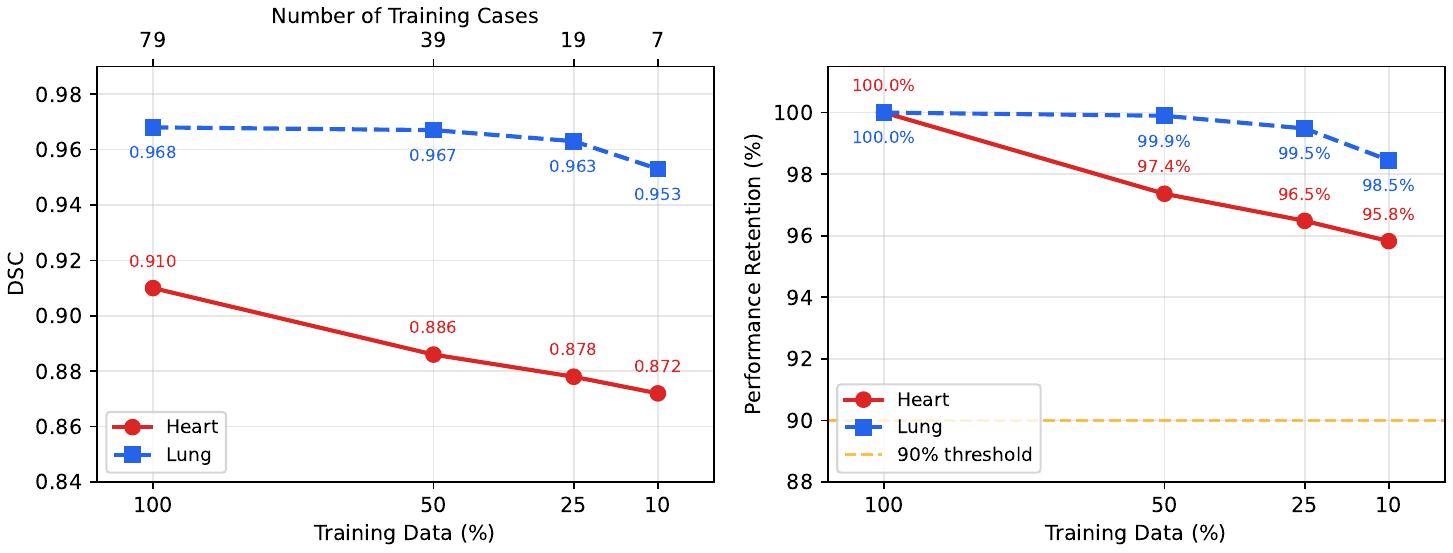}
\caption{Data efficiency: DSC (left) and performance retention (right) as training set size decreases. Both organs retain $>$95\% of full-data accuracy with only 7 cases.}
\label{fig:ablation_data}
\end{figure}

\paragraph{Effect of Training Set Size.}
We trained four models using 79, 39, 19, and 7 annotated 3D CT volumes (100\%, 50\%, 25\%, and 10\%) and evaluated each on the held-out test set.
The results are summarized in Table~\ref{tab:ablation_data} and Fig.~\ref{fig:ablation_data}.
For the targets prompted with "heart" and "lung", the DSC gap between the largest (79 cases) and smallest (7 cases) training sets remains under 0.04, demonstrating that the proposed adaptation (LoRA + HNM) reaches near-saturation performance with remarkably few samples.
This confirms that LoRA's low-rank update preserves the pretrained representations, requiring only a few domain-specific examples to overcome the natural-to-medical distribution shift.

\begin{table}[t]
\centering
\caption{Data-efficiency ablation. $\Delta$DSC denotes the drop relative to the 100\% baseline.}
\label{tab:ablation_data}
\begin{tabular}{r cc cc}
\toprule
 & \multicolumn{2}{c}{\textbf{``heart''}} & \multicolumn{2}{c}{\textbf{``lung''}} \\
\cmidrule(lr){2-3} \cmidrule(lr){4-5}
\textbf{Training Cases (\%)} & DSC & $\Delta$DSC & DSC & $\Delta$DSC \\
\midrule
79 (100\%) & 0.910 & --    & 0.968 & --    \\
39 (50\%)  & 0.886 & 0.024 & 0.967 & 0.001 \\
19 (25\%)  & 0.878 & 0.032 & 0.963 & 0.005 \\
7 (10\%)   & 0.872 & 0.038 & 0.953 & 0.015 \\
\bottomrule
\end{tabular}
\end{table}

\begin{figure}[t]
\centering
\includegraphics[width=\columnwidth]{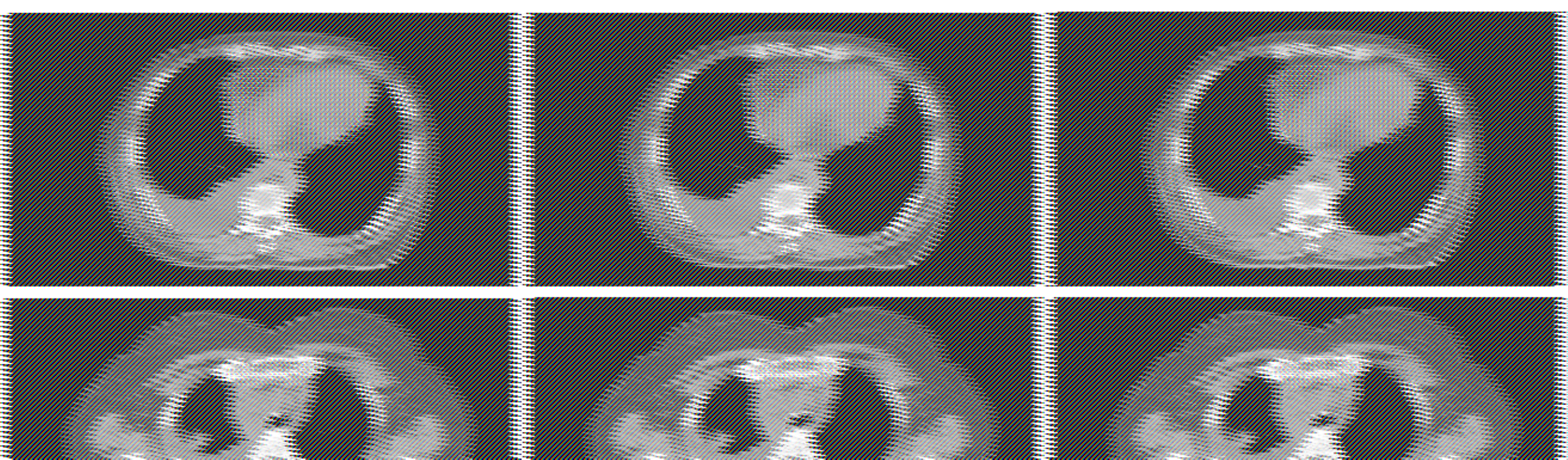}
\caption{Spatiotemporal filtering on representative cardiac axial slices. \textbf{(a)}~AIP (average of 10 phases). \textbf{(b)}~Union ITV (red). \textbf{(c)}~Refined ITV (green, $R_{\min}{=}3$). \textbf{(d)}~Retained ITV (green) and removed noise (yellow). Top: cardiac boundary; bottom: distant slice.}
\label{fig:itv_qualitative}
\end{figure}

\paragraph{4DCT ITV Generation.}
To evaluate clinical applicability, we applied the trained model (LoRA~+~HNM) to the TCIA 4DCT dataset (20 patients, 10 phases each) without further fine-tuning, filtering predictions with $R_{\min}{=}3$ and $V_{\min}{=}100$ voxels.
Since no expert-delineated ITV ground truth is available, the generated contours were visually assessed by a radiation oncologist.
Across all 20 patients, the ITVs were judged anatomically plausible, suggesting that a model trained solely on 3D CT generalizes to 4DCT phases and that spatiotemporal filtering effectively integrates phase-wise predictions into a clinically usable ITV.
The effect of spatiotemporal filtering on representative cardiac axial slices is illustrated in Fig.~\ref{fig:itv_qualitative}.

\section{Discussion and Conclusion}

We have presented a data-efficient, resource-aware framework for automated ITV generation from 4DCT.
By adapting SAM~3 through LoRA on only seven annotated 3D CT volumes---with a DSC degradation of under 0.04 relative to 79-case training---and applying it directly to unseen 4DCT phases, our approach circumvents the two most cited barriers to clinical deployment of deep learning segmentation: data scarcity and computational cost.
The compact \SI{3.2}{GB} model trained on a single consumer-grade GPU (\SI{10}{GB} VRAM) is directly transferable to clinical workstations, eliminating the dependency on dedicated computing infrastructure.

Incorporating the HNM strategy substantially reduces HD95 by eliminating distant false positives, and spatiotemporal filtering further suppresses sporadic artifacts at the ITV level.
Visual assessment by a radiation oncologist confirmed anatomically plausible ITVs across all 20 TCIA patients, despite the absence of 4DCT-specific training data---an important clinical benefit, as spurious contour regions inflate the ITV and potentially expose healthy tissue to unnecessary dose.

From an adaptive radiotherapy perspective~\cite{yan1997adaptive}, the pipeline's efficiency facilitates rapid re-segmentation within minutes upon the acquisition of new 4DCT scans, and the text-prompted interface allows clinicians to target different structures without retraining.

The primary limitations of this study are twofold: (1)~the 4DCT evaluation relies on qualitative expert assessment due to the inherent clinical difficulty of establishing voxel-perfect ITV ground truth across continuous respiratory phases, and (2)~the dataset originates from a single institution and scanner vendor (TCIA CT-vs-PET-Ventilation-Imaging, 20 patients)~\cite{https://doi.org/10.7937/3ppx-7s22}.
Multi-center validation with quantitative evaluation against consensus expert contours would strengthen the generalizability claim.
Additionally, extending this zero-shot framework to gross tumor volume (GTV) segmentation in the presence of large respiratory excursion (e.g., lower-lobe lung tumors) warrants dedicated investigation, as tumor boundaries are inherently more ambiguous than those of organs at risk (OARs) such as the heart and lungs.

In conclusion, our results demonstrate that combining parameter-efficient foundation model adaptation with spatiotemporal 4DCT integration provides a scalable, highly data-efficient proof-of-concept for automated motion envelope generation. Validated on principal thoracic organs, this framework demonstrates that robust 4DCT ITV generation is achievable even without 4DCT-specific training data, establishing a strong foundation for future extension to complex tumor ITVs.

%% Disclosure of Interest (restore for camera-ready version):
%\begin{credits}
%\subsubsection{\discintname}
%J.S. Kim is the CEO and a shareholder of Oncosoft Inc. S. Moon is the Chief Product Officer (CPO) and a shareholder of Oncosoft Inc. C. Song, and W. Cho are employees of Oncosoft Inc. The remaining authors have no competing interests to declare.
%\end{credits}

%
\newpage
\bibliographystyle{splncs04}
\bibliography{references}
\end{document}